\documentclass[a4paper,10pt,twoside]{cpc-hepnp}
\usepackage{CJK,upgreek,fancyhdr}
\usepackage{multicol}
\usepackage{graphicx}
\usepackage{booktabs}
\usepackage{amssymb,bm,mathrsfs,bbm,amscd}
\usepackage[tbtags]{amsmath}
\usepackage{lastpage}

\usepackage{xcolor}

\begin{document}
	
%	\fancyhead[c]{\small Chinese Physics C~~~Vol. xx, No. x (201x) xxxxxx}
	\fancyfoot[C]{\small 010201-\thepage}

	\title{Simple Woods-Saxon type form 
		for $\Omega \alpha $ and $\Xi \alpha$ interactions using Folding Model}
	
	\author{%
		Faisal Etminan$^{1}$\email{fetminan@birjand.ac.ir}%
		\quad Mohammad Mehdi Firoozabadi$^{1}$\email{mfiroozabadi@birjand.ac.ir}%
	}
	\maketitle

	\address{%
		$^1$ Department of Physics, Faculty of Sciences, University of Birjand, Birjand 97175-615, Iran\\
	}

	\begin{abstract}
		We derive a simple Woods-Saxon type form for the potentials between $Y=\Xi, \Omega$ and $\alpha$ by using a single-folding potential method, based on a separable $Y$-nucleon potential. Accordingly, the potentials $\Xi+\alpha$ and $\Omega+\alpha$ are obtained using the ESC08c Nijmegens $\Xi N$ potential
		(in $^{3}S_{1}$ channel) and HAL QCD Collaboration $\Omega N$ interactions
		(in lattice QCD), respectively. In deriving the  potential between $Y$ and $\alpha$, the same potential between $Y$ and $N$ is used. Binding energy, scattering length and effective range of $Y$ particle on the alpha particle are approximated by the resultant potentials. The depths of the potentials in $\Omega \alpha $ and $\Xi \alpha $ systems are obtained $-61$ and $-24.4$ MeV, respectively. In the case of $\Xi \alpha$ potential, a fairly good agreement is observed between the single-folding potential method and the phenomenological potential of Dover-Gal model. These potentials can be used in 3-,4- and 5-body cluster structures of $ \Omega$  and $\Xi$ hypernuclei. 
	\end{abstract}


	\begin{multicols}{2}
		\section{Introduction} \label{sec:intro}
		The  $\Xi$ and $ \Omega$ hypernuclei offer a mixture of simplicity
		and fundamental interest in connection with the interactions between
		nucleons and strange particles. Hyperon interactions are not known sufficiently well due to the limited scattering data. However, more precise data on light hypernuclei (from high resolution gamma-ray experiments~\citep{Tamura2000}), advanced few-body theoretical methods~\cite{hiyama2000,hiyama2008,hiyama2014,hiyama2019,Moosavi2019,Sekihara2018,Garcilazo2018,Garcilazo2019}, quark delocalization, color screening~\cite{Ping199}, constituent quark models~\citep{huang2015}, lattice QCD calculations~\cite{etminanmpla2014,nemura2014,Sasaki2018} and femtoscopic analyses of pp, pA and AA collisions in the ALICE and STAR experiments~\citep{Acharya2019} have recently been providing us with valuable information. The KEK-E373 experiment reported the first evidence of a bound $\Xi$-hypernucleus $^{14}N+\Xi$, i.e., the so called KISO event~\citep{Nakazawa2015}. Recently, the first experimental observation of an attractive strong interaction between a proton and a hyperon $ \Xi $ has been reported by ALICE collaboration~\citep{Acharya2019}.

		The theoretical efforts of the lattice HAL QCD Collaboration have led to the derivation of baryon-baryon interactions near the physical pion mass~\citep{aoki2012}. The most recent results of these efforts hint to the existence of shallow bound states in $\Omega N$ systems~\citep{Iritani2019prb}. We study the $\Omega+\alpha$ system making use of this $\Omega N$ potential. The depth of the potential in $\Omega \alpha $ and $\Xi \alpha$ systems are uncertain, because there are not enough experimental data for their nuclear bound states. However, for the latter, some diverse phenomenological potentials are considered~\cite{dover1983,Filikhin2008,Garcilazo2016}. 
		
		Motivated by the aforementioned description, and vast applications of $\Omega \alpha $ and $\Xi \alpha$ interactions in 3-,4- and 5-body cluster structures of $ \Omega$ hypernuclei (describing the response of 3$\alpha$ system to the addition of $\Omega$ particle)~\citep{hiyama2000}, $\alpha$ cluster model approach~\citep{hiyama2014} and variational four-body calculation~\citep{Garcilazo2020}, we present $\Omega \alpha $ and $\Xi \alpha$ interactions in a simple Woods-Saxon type form.

		In order to test and validate our method, we apply it to $\Xi \alpha$ systems and compare the results with the phenomenological Dover-Gal (DG) potential type~\citep{dover1983}. 
		Additionally, we use ESC08c Nijmegen model $\Xi N$ potential in $^{3}S_{1}$ channel~\citep{Nagels}. 
		
		Here, we consider a $Y +\alpha$ system because of the low compressibility of the $\alpha$-cluster and high reaction thresholds that enable us to employ a one-channel approximation over a wide energy range. The $Y+\alpha$ system is studied using a single-folding potential (SFP) method. In this model, the $Y+\alpha$ system consists of an alpha and a $ Y $ particle moving in the effective $Y \alpha$ potential. The effective nuclear potential is approximated by the single-folding of nucleon density $\rho\left(\vec{r}^{\prime}\right)$ in
		the $\alpha$-particle and hyperon-nucleon potential $V_{YN}\left(\left|\vec{r}-\vec{r}^{\prime}\right|\right)$ between the $Y$ particle at $\vec{r}^{\prime}$
		and the nucleon at $\vec{r}$~\cite{Satchler1979,Miyamoto2018}. Then, the resulting
		$Y\alpha$ potential is fitted to a separable form. Finally, we
		solve the Schr\"{o}dinger equation by using the fitted $Y \alpha$
		potential in the infinite volume and extract its scattering observables
		from the asymptotic behavior of the wave function. The model is expected
		to be accurate only for the low energy properties of the $Y+\alpha$ system
		since it is based on a $Y N$ potential, which is fitted to low
		energy $Y N$ scattering parameters.
		
		We should emphasize that the coupling of $ \Omega N $ to higher-mass ($ \Lambda\Xi^{*} $) and lower-mass ($ \Lambda\Xi $ and $ \Sigma\Xi $) channels is not taken into account since we assume that these contributions are of second order of smallness to the binding energy of few-body systems~\citep{Iritani2019prb}. To draw a definite conclusion about the binding energy of $\Omega +\alpha$ system, it is necessary to perform a coupled-channel analysis of the HAL QCD method~\citep{aoki2013}. Moreover, in the calculations, the Coulomb force is not taken into account.

		This paper is organized as follows. In Sec.~\textnormal{\ref{sec:folding-Model}}, after a 
		brief discussion on the single-folding potential method, we introduce and parametrize all dominating set of input parameters, i.e., those of the two-body %
			potentials. In Sec.~\textnormal{\ref{sec:result}}, we present and discuss the results. Finally, in Sec.~\textnormal{\ref{sec:summery-conclusion}}, a summary and conclusions are presented.

		\section{Single-folding potential model} \label{sec:folding-Model}
		
		We obtain the effective potential of $Y+\alpha$ systems by
		using the single-folding potential model~\cite{Satchler1979,Miyamoto2018}. This method is briefly outlined in the following. The $Y \alpha$ potential is defined as
		
		\begin{equation}
		V_{Y\alpha}\left(\vec{r}\right)=\int\rho\left(\vec{r}^{\prime}\right)V_{Y N}\left(\left|\vec{r}-\vec{r}^{\prime}\right|\right)d\tau^{\prime},\label{eq:V_alfaOmega}
		\end{equation}
		
		where $\rho\left(\vec{r}^{\prime}\right)$ is the nucleon density in
		the $\alpha$-particle at a distance $\vec{r}^{\prime}$ from its
		center-of-mass, and is given by~\citep{Akaishi1986},
		
		\begin{equation}
		\rho\left(\vec{r}^{\prime}\right)=4\left(\frac{4\beta}{3\pi}\right)^{3/2}\exp\left(-\frac{4}{3}\beta r^{\prime2}\right).\label{eq:nucleon-density}
		\end{equation}
		
		The integration in Eq.~(\ref{eq:V_alfaOmega}) is over all space as permitted
		by $\rho\left(\vec{r}^{\prime}\right)$. The required normalization
		condition is satisfied by 
		
		\begin{equation}
		\int\rho\left(\vec{r}^{\prime}\right)d\tau^{\prime}=4\left(\frac{4\beta}{3\pi}\right)^{3/2}\intop_{0}^{\infty}\mathop{\exp\left(-\frac{4}{3}\beta r^{\prime2}\right)4\pi r^{\prime2}dr^{\prime}}=4.
		\end{equation}
		
		The constant $\beta$ is determined from the rms radius of $\textrm{\ensuremath{^{4}}He}$~\citep{Akaishi1986}, 
		
		\begin{equation}
		\textrm{\ensuremath{r_{r.m.s}}}=\frac{3}{\sqrt{8\beta}}=1.47\:\textrm{fm}.
		\end{equation}
		
		In Eq.~(\ref{eq:V_alfaOmega}), $V_{YN}\left(\left|\vec{r}-\vec{r}^{\prime}\right|\right)$ is the potential in configuration space between the $Y$ particle at $\vec{r}$
		and the nucleon at $\vec{r}^{\prime}$.
		
		We take the $\Xi N$ potential
		in $^{3}S_{1}$ channel and simulate the ESC08c Nijmegen model, which consists
		of local central Yukawa-type potentials with attractive and repulsive terms~\cite{Nagels,Garcilazo2016},
		
		\begin{equation}
		V_{\Xi N}\left(\vec{r}\right)=-568\frac{\exp\left(-4.56r\right)}{r}+425\frac{\exp\left(-6.73r\right)}{r}. \label{eq:pot-XiN}
		\end{equation}
		
		The low-energy data of this potential is listed in Table~\ref{tab:ERE-ON-XiN}. In the case of $\Omega N$, we use S-wave and spin $2$ $\Omega N$
		potential, which is given by HAL QCD Collaboration with nearly physical
		quark masses~\citep{Iritani2019prb}. The lattice discrete potential is fitted by an analytic function composed of an attractive Gaussian core plus a long range
		\textrm{(Yukawa\ensuremath{)^{2}}} attraction with a form factor of~\citep{etminan2014},
		
		\begin{equation}
		V_{\Omega N}\left(r\right)=b_{1}e^{\left(-b_{2}r^{2}\right)}+b_{3}\left(1-e^{-b_{4}r^{2}}\right)\left(\frac{e^{-m_{\pi}r}}{r}\right)^{2},\label{eq:NOmega_pot}
		\end{equation}
		the pion mass in Eq.~(\ref{eq:NOmega_pot}), which is taken from the simulation, is $\textrm{\ensuremath{m_{\pi}}=146}$ MeV. The lattice results are fitted reasonably well, $\chi^{2}/d.o.f\simeq1$, with four different sets of parameters given in Table \ref{tab:Fit_para}. 
		The low-energy data of this potential is also given in Table~\ref{tab:ERE-ON-XiN}.
		
		\begin{center}
			\tabcaption{ The low-energy parameters, scattering length, $a_{0}$, effective range, $r_{0}$,  and binding energy, $B_{Y N}$, of ESC08c Nijmegen $\Xi N$~\cite{Nagels,Garcilazo2016} given by Eq.~(\ref{eq:pot-XiN}), and the HAL QCD $\Omega N$ potential~\citep{Iritani2019prb} given by Eq.~(\ref{eq:NOmega_pot}). \label{tab:ERE-ON-XiN}}
			\footnotesize
			\begin{tabular*}{80mm}{c@{\extracolsep{\fill}}cccc}
				\toprule	System &Channel& $a_{0}$(fm)&$r_{0}$(fm)& $B_{Y N}$(MeV) \\
				\hline 		
				$\Omega N$ & $ ^5S_{2}$ & 5.30 & 1.26  & 1.54 \\
				$\Xi    N$ & $^3S_{1}$  & 4.91 & 0.527 & 1.67   \\
				\bottomrule
			\end{tabular*}
		\end{center}

		\begin{center}
			\tabcaption{Fitting parameters in Eq.~(\ref{eq:NOmega_pot}) for different models, $P_{i}$, for $^{5}S_{2}$ $\Omega N$ interaction~\citep{Iritani2019prb}.\label{tab:Fit_para} }
			\footnotesize
			\begin{tabular*}{80mm}{c@{\extracolsep{\fill}}cccc}
				\toprule	& $P_{1}$ & $P_{2}$ & $P_{3}$ & $P_{4}$\\
				\hline  
				$b_{1}\left(\mathrm{MeV}
				\right)$ & -306.5  & -313.0 & -316.7 & -296\\
				$b_{2}\left(\mathrm{fm}^{-2}\right)$ & 73.9  & 81.7 & 81.9  & 64\\
				$b_{3}\left(\mathrm{MeV}.fm^{-2}\right)$ & -266  &  -252 & -237 & -272\\
				$b_{4}\left(\mathrm{fm}^{-2}\right)$ & 0.78  & 0.85  & 0.91  & 0.76\\
				\bottomrule
			\end{tabular*}
		\end{center}
		
		\section{Results} \label{sec:result}
		
		In order to test $\Xi \alpha$ potential obtained from SFP model,
		we employed the phenomenological potential of Woods-Saxon type for
		the $\Xi \alpha$ interaction by using Dover-Gal (DG) model given
		in~\cite{Filikhin2008,dover1983} 
		
		\begin{equation}
		V_{\Xi\alpha}^{DG}\left(r\right)=-V_{0}\left[1+\exp\left(\frac{r-R}{c}\right)\right]^{-1} , \label{eq:DG}
		\end{equation}
		
		where $ V_{0}$ is the depth parameter, $ R=1.1A^{\left(1/3\right)}$ with $ A $ being the mass number of the nuclear core (here, $A=4$ for the alpha particle) and $c$ is the surface diffuseness. The values of these three parameters in DG model are given in Table~\ref{tab:ERE-DO-SFP}. The $\Xi+\alpha$ system is bound by this potential with an energy
		of $E_{B}=-2.1$ MeV. It is important to note that DG potential has no repulsive core.
		
	\end{multicols}
	
	\begin{center}
		\tabcaption{DG potential model parameters of Eq.~(\ref{eq:DG}) from~\cite{Filikhin2008,dover1983}. The values of fitting parameters of $\Xi \alpha$ are obtained by fitting SFP model to a function in the same form as Eq.~(\ref{eq:DG}). The corresponding low-energy parameters, scattering length, effective range and binding energy of both models are given. The obtained results, using experimental masses of $\alpha$ and $\Xi$, are $3727.38\:\textrm{MeV}/c^{2}$ and $1318.07\:\textrm{MeV}/c^{2}$, respectively.\label{tab:ERE-DO-SFP}}. 
		\footnotesize	
		\begin{tabular*}{95mm}{ccccccc}
			\toprule			Model & $ V_{0} $ (MeV)& $ R $(fm)& $ c $(fm)& $a_{0}$(fm)&$r_{0}$(fm)& $B_{\Xi \alpha}$(MeV) \\
			\hline 		
			DG  & 24   & 1.74 & 0.65 & -4.9 & 1.9 & -2.1\\
			SFP & 24.4 & 1.72 & 0.31 & -6.6 & 1.9 & -1.54\\
			\bottomrule
		\end{tabular*}
	\end{center}
	
	\begin{multicols}{2}
		
		To obtain observables, such as scattering phase shifts and binding
		energy, we fit $V_{Y\alpha}\left(\vec{r}\right)$ to the Wood-Saxon form using the function in Eq.~(\ref{eq:DG}) with three parameters $ V_{0}, R$ and $c$. The results of fitting for these parameters are presented in Table~\ref{tab:ERE-DO-SFP}.
		
		The obtained single-folding potential, $V_{\Xi\alpha}\left(\vec{r}\right)$, its corresponding fit function, $ V_{fit}(r) $, and  $V_{\Xi\alpha}^{DG}\left(r\right)$ (for comparison) are shown in Fig.~\ref{fig:vc-Xi-fit-DG}. We solve the Schr\"{o}dinger equation with the fitted potential in the infinite volume and extract its scattering observables from the asymptotic behavior of the wave function. For comparison, in Fig.~\ref{fig:phase-xialfa-DG-SFP}, the phase shifts from DG and SFP model potentials are shown.
		
		\begin{center}
			\includegraphics[width=8cm]{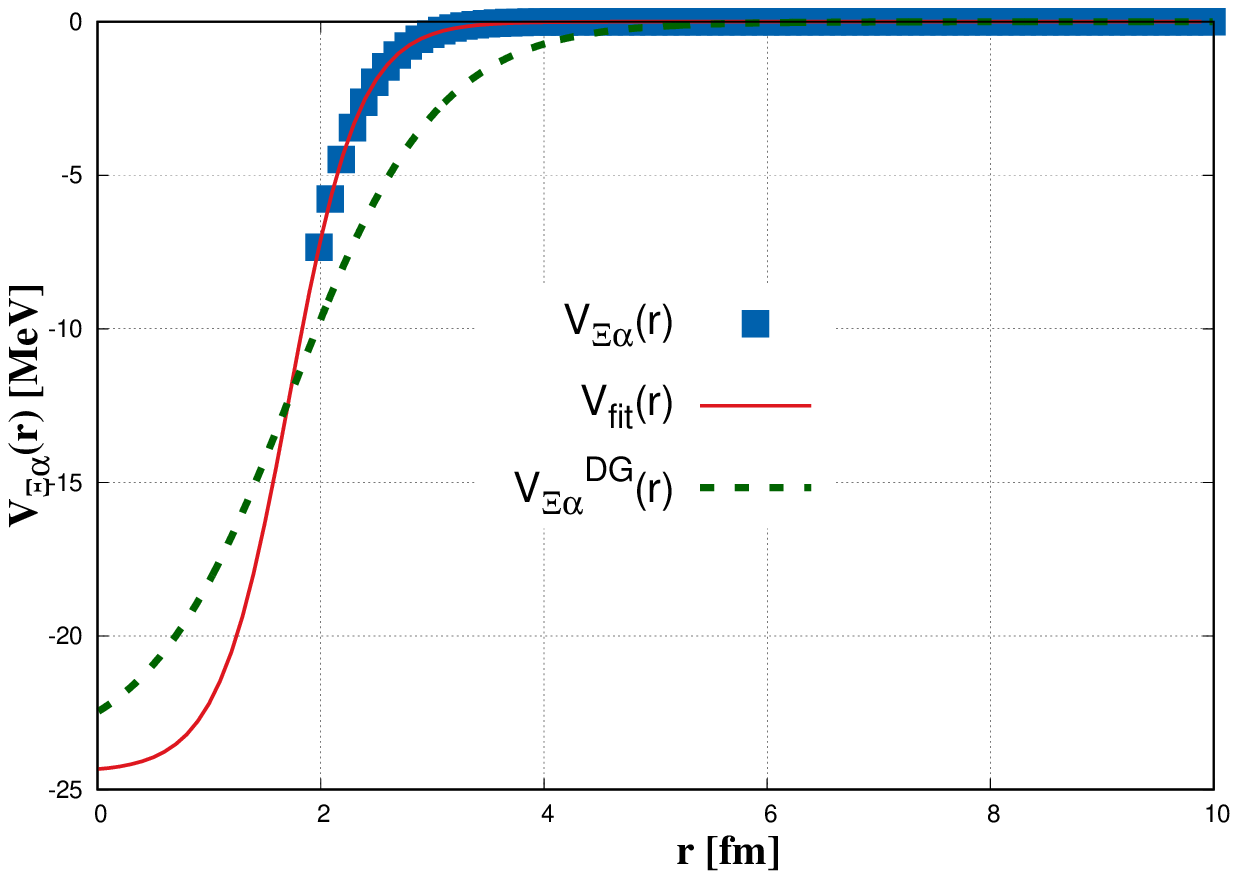}
			\figcaption{The single-folding potential, $V_{\Xi\alpha}\left(\vec{r}\right)$, for the $\Xi N$  interaction in $^{3}S_{1}$ channel given in~\cite{Nagels,Garcilazo2016}. $V_{fit}(r)$ (Red line) shows the results of the fitting by using the same form as Eq.~(\ref{eq:DG}). For comparison, we also present Dover-Gal $\Xi\alpha$ potential (green dashed line)~\cite{Filikhin2008,dover1983}, i.e.,  $V_{\Xi\alpha}^{DG}\left(r\right)$ in Eq.~(\ref{eq:DG}).  
				\label{fig:vc-Xi-fit-DG}}
		\end{center}

		\begin{center}
			\includegraphics[width=8cm]{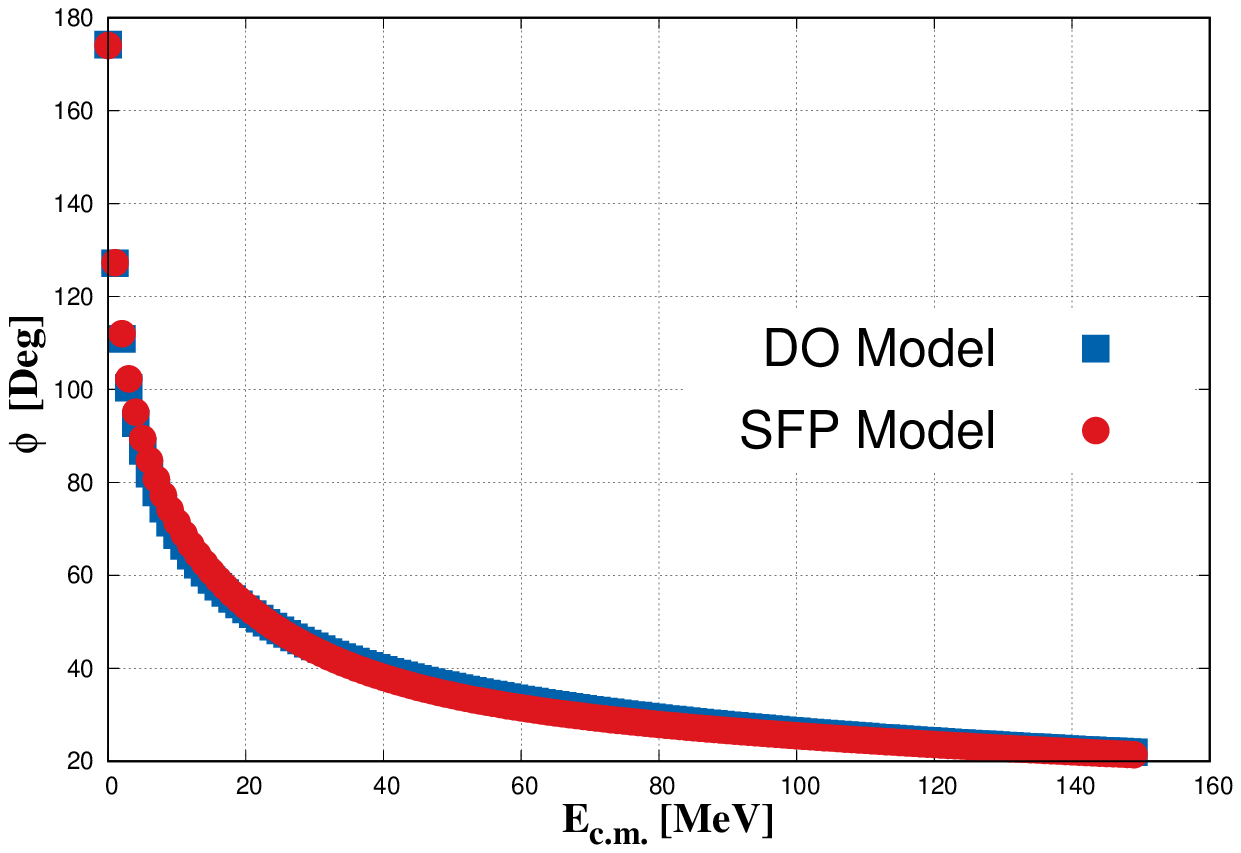}
			\figcaption{ The extracted $\Xi\alpha$ phase shifts for two potential models, DG and SFP, are given in Fig.~\ref{fig:vc-Xi-fit-DG} for comparison. According to this figure, one can see a fairly good agreement between these two models. \label{fig:phase-xialfa-DG-SFP}}
		\end{center}
		
		The effective range expansion (ERE) of the phase shifts up to the
		next-leading-order (NLO) reads
		
		\begin{equation}
		k\cot\delta_{0}=-\frac{1}{a_{0}}+\frac{1}{2}r_{0}k^{2}+\mathcal{O}\left(k^{4}\right),\label{eq:ERE}
		\end{equation}
		
		with $\left(a_{0}\right)$ and $\left( r_{0}\right)$ being the scattering
		length and effective range, respectively. The results of the calculations for the binding energy and the ERE parameters $\left(a_{0}, r_{0}\right)$ are given in Table~\ref{tab:ERE-DO-SFP}. According to the results 
		in Table~\ref{tab:ERE-DO-SFP}, a good agreement can be observed between the single-folding potential method and the phenomenological potential of Dover-Gal model.

		The single-folding potential $V_{\Omega\alpha}\left(\vec{r}\right)$
		for different models of $\Omega N$ interaction ($ P_{i} $, Table~\ref{tab:Fit_para}) are shown in Fig.~\ref{fig:vcF} (a). We summarize the results of fitting and the corresponding parameters in Fig.~\ref{fig:vcF} (b) and Table~\ref{tab:Fit_para-of-Oalfa-pot}, respectively. We adjust the depth $ V_{0} $ in such a way to produce the best fit for $R$ and $c$ parameters, i.e., $\chi^{2}/d.o.f\simeq1 $.

			Garcilazo and Valcarce~\cite{Garcilazo2018} showed that $\Omega N$  $^{5}S_{2}$ and $NN$ $^{3}S_{1}$ channels give rise to a $\Omega d$ bound state in the state with maximal spin $\left(I,J^{P}\right)=\left(0,5/2^{+}\right)$ with a binding energy of $\sim17$ MeV measured with respect to the $NN\Omega$ threshold by solving the three-body bound-state Faddeev equations. Here, we obtain an $\Omega\alpha$ binding energy of $\sim23$ MeV, as shown in Table~\ref{tab:ERE-oalfa}. Since the potential must be more attractive than the approximated single-folding potential, the resulting energy is only an upper bound for the $\Omega \alpha$ system, and the binding energy of $\Omega\alpha$ is greater than of $\Omega d$, which seems completely reasonable.

		\begin{center}
			\tabcaption{
				The values of parameters are obtained by fitting the $\Omega \alpha$ SFP potential to a function in the same form as Eq.~(\ref{eq:DG}), for different models, $P_{i}$ , of the $\Omega N$ interaction~\citep{Iritani2019prb}.\label{tab:Fit_para-of-Oalfa-pot}}
			\footnotesize
			\begin{tabular*}{55mm}{ccccc}
				\toprule
				& $P_{1}$ & $P_{2}$ & $P_{3}$ & $P_{4}$\\
				\hline 
				$V_{0}\left(\textrm{MeV}\right)$ &-61  & -61  & -61  & -61\\
				$R\left(\textrm{fm}\right)$  & 0.47 & 0.47  & 0.47 & 0.47\\
				$c\left(\textrm{\textrm{fm}}\right)$ & 1.7 & 1.7 & 1.7 & 1.7 \\
				\bottomrule
			\end{tabular*}
		\end{center}

		\begin{center}
			\includegraphics[width=8cm]{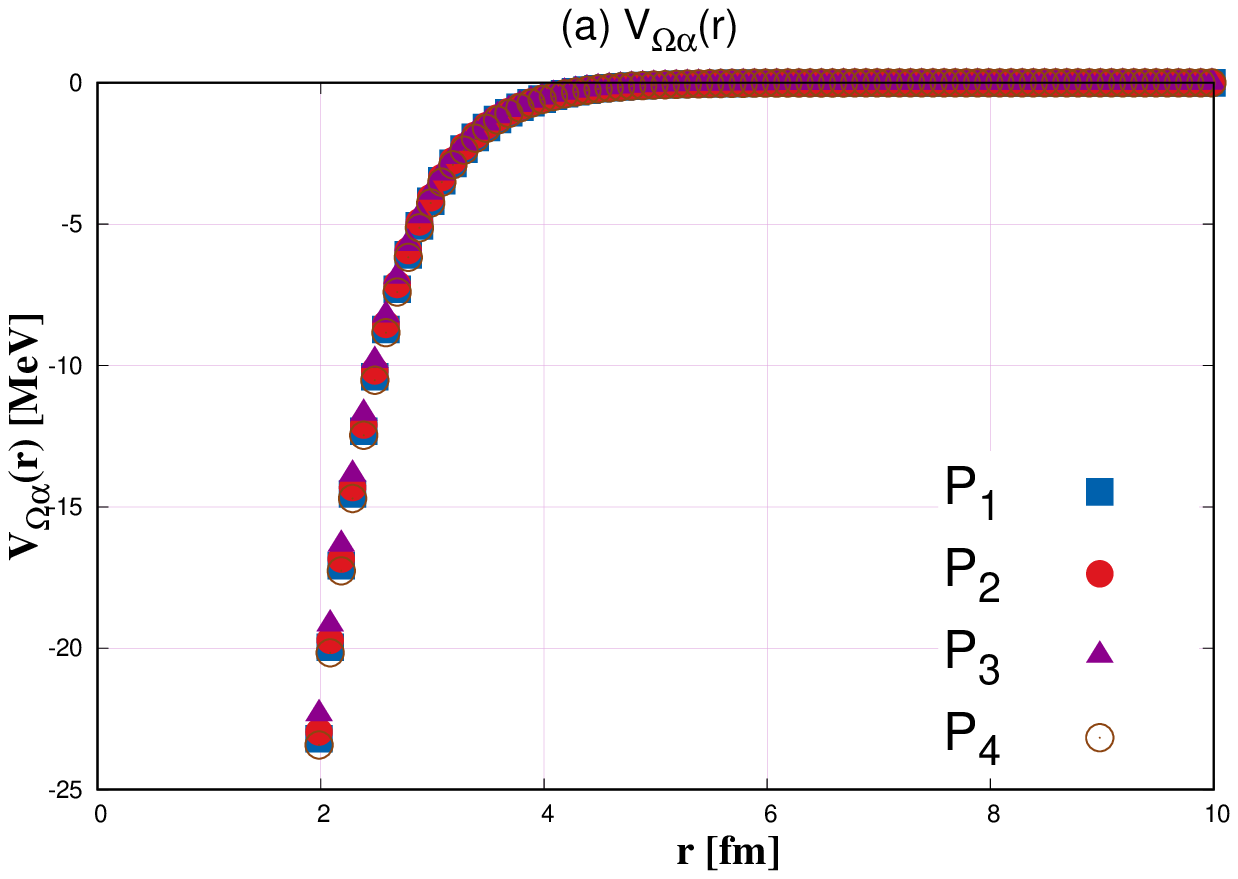}
			\includegraphics[width=8cm]{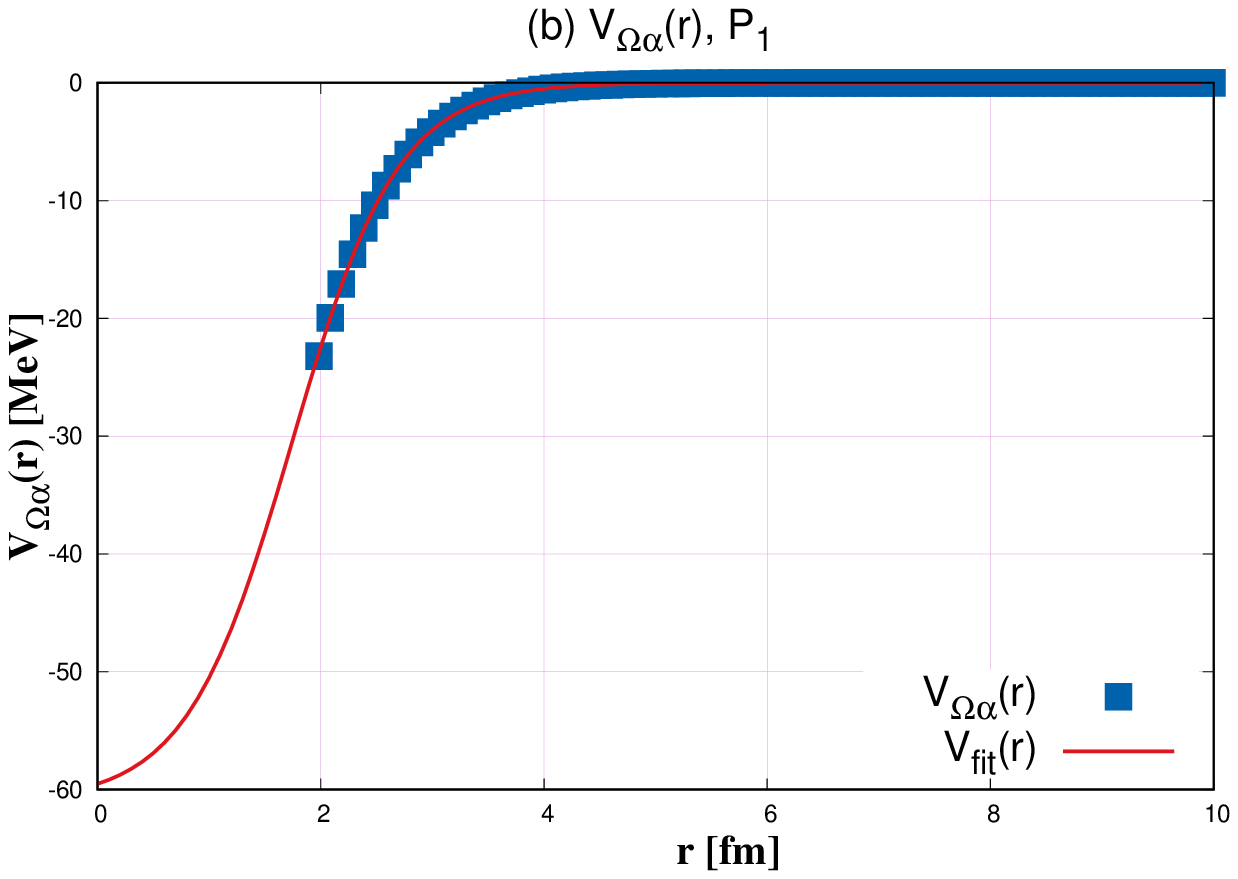} 
			\figcaption{(a) The single-folding potential, $V_{\Omega\alpha}\left(\vec{r}\right)$,
				for different models of $\Omega N$ interaction ($ P_{i} $) given in Table~
				\ref{tab:Fit_para}. (b) The single-folding potential, $V_{\Omega\alpha}\left(\vec{r}\right)$,
				and $V_{fit}(r)$ fit function (Red line), which has the same form as Eq.~(\ref{eq:DG}) for the set $P_{1}$. \label{fig:vcF}}
		\end{center}
		
		Shown in Fig.~\ref{fig:Phase-shift.} is the S-wave scattering phase shift $\delta_{0}$ as a function of the kinetic energy. In Table~\ref{tab:ERE-oalfa}, we present the binding energies and ERE parameters $\left(a_{0},r_{0}\right)$ obtained from $\Omega \alpha$ phase shifts, for different models of $\Omega N$ interaction reported in
		Ref.~\citep{Iritani2019prb} and summarized in Table~\ref{tab:Fit_para}.
		
		\begin{center}
			\includegraphics[width=8cm]{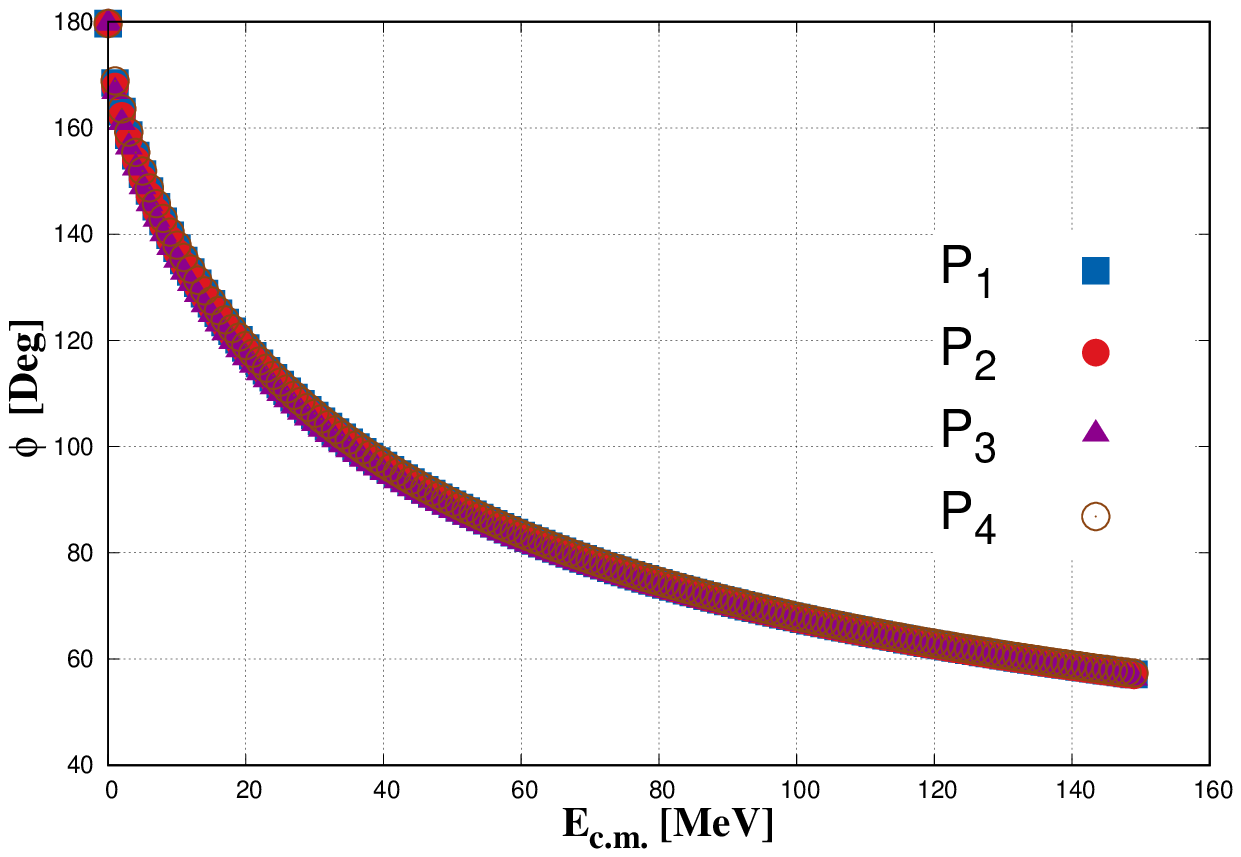}
			\figcaption{The S-wave scattering phase shift $\delta_{0}$ as a function of the kinetic energy, $k^{2}/(2\mu)$.\label{fig:Phase-shift.}}
		\end{center}
	\end{multicols}

	\begin{center}
		\tabcaption{Scattering length, $a_{0}$, effective range, $r_{0}$, and binding energy, $B_{\Omega\alpha}$, of $\Omega\alpha$ for the different models of $\Omega N$ interaction given in Table~\ref{tab:Fit_para}~\citep{Iritani2019prb}. The results that are obtained using the experimental masses of $\alpha$ and $\Omega$ are $3727.38\:\textrm{MeV}/c^{2}$ and $1672.45\:\textrm{MeV}/c^{2}$, respectively. The values in parentheses correspond to the masses of $\alpha$ and $\Omega$ derived by the HAL QCD Collaboration, i.e., $3818.8\:\textrm{MeV}/c^{2}$and $1711.5\:\textrm{MeV}/c^{2}$, respectively~\citep{Iritani2019prb}.\label{tab:ERE-oalfa}}
		\footnotesize	
		\begin{tabular*}{100mm}{@{\extracolsep{\fill}}ccccc}
			\toprule
			& $P_{1}$ & $P_{2}$ & $P_{3}$ & $P_{4}$\\
			\hline 
			\footnotesize		
			$a_{0}$ (fm) &-0.99(-0.93)& -1.01(-0.96) & -1.05(-1.00) & -0.98(-0.92)\\
			$r_{0}$ (fm) & 0.67(0.67) & 0.67(0.67) & 0.67(0.67) & 0.67(0.68) \\
			$B_{\Omega\alpha}$ (MeV) & -22.9 (-23.3) & -22.8(-23.2) & -22.4(-22.8) & -23.0(-23.4)\\
			\bottomrule
		\end{tabular*}
	\end{center}
	
	\begin{multicols}{2}

		\section{Summary and conclusions} \label{sec:summery-conclusion}
		
		We derived a simple Woods-Saxon type form for the potentials of $\Xi+\alpha$ and $\Omega+\alpha$ systems by making use of, respectively, the ESC08c Nijmegen $\Xi N$ potential in $^{3}S_{1}$ channel and the HAL QCD Collaboration $\Omega N$  potential in $^{5}S_{2}$ channel plus density function of the alpha particle in single-folding potential method. 
		
			We showed that the effective central folding potential of $\Omega \alpha$ may have simple Wood-Saxon form and estimated the upper bound for binding energy of $\Omega$ particle on a $\alpha$.
		
		Our method was tested against the phenomenological potential of Woods-Saxon type for the $\Xi\alpha$ interaction by the Dover-Gal model and a fairly good agreement was found between the two methods.
		
		The scattering length and the effective range were obtained by solving the
		Schr\"{o}dinger equation using the resultant potential. The binding energies
		of $\Xi+\alpha$ and $\Omega+\alpha$ systems were about $-1.5$ and $-23$ MeV, respectively. These results indicate that $\Omega\alpha$ hypernuclei are deeply bound states or resonances, which may be experimentally observed in the real world. 
		
		We should emphasize that the cases of coupling of $ Y N $ to higher- and lower-mass channels were not taken into account. The calculations do not also take into account the Coulomb force. To draw a definite conclusion about the binding energy of $Y  \alpha$, it is necessary to do a coupled-channel analysis. 
		
		We hope that our results can be used as tests of various theoretical
		models for the exotic nuclei structures, especially in few $\alpha$ cluster structures of $ \Omega$ hypernuclei (describing the response of the few $\alpha$ systems to the addition of the $\Omega$ particle)~\citep{hiyama2000}, $\alpha$ cluster model approach~\citep{hiyama2014} and possible future experiments, where these lattice-QCD-based predictions may be tested.
		\vspace{3mm}
		
		%\bibliography{Oalfa_ver03}
		
	\end{multicols}

	\clearpage
\end{document}